# Long-term perturbations due to a disturbing body in elliptic inclined orbit

Xiaodong Liu[1], Hexi Baoyin[2], and Xingrui Ma[3]

*School of Aerospace, Tsinghua University, 100084 Beijing, China*

Email:   liu-xd08@mails.tsinghua.edu.cn; baoyin@tsinghua.edu.cn;

maxr@spacechina.com

**Abstract**

In the current study, a double-averaged analytical model including the action of the perturbing body's inclination is developed to study third-body perturbations. The disturbing function is expanded in the form of Legendre polynomials truncated up to the second-order term, and then is averaged over the periods of the spacecraft and the perturbing body. The efficiency of the double-averaged algorithm is verified with the full elliptic restricted three-body model. Comparisons with the previous study for a lunar satellite perturbed by Earth are presented to measure the effect of the perturbing body's inclination, and illustrate that the lunar obliquity with the value 6.68º is important for the mean motion of a lunar satellite. The application to the Mars-Sun system is shown to prove the validity of the double-averaged model. It can be seen that the algorithm is effective to predict the long-term behavior of a high-altitude Martian spacecraft perturbed by Sun. The double-averaged model presented in this paper is also applicable to other celestial systems.

---

[1]  PhD candidate, School of Aerospace, Tsinghua University

[2]  Associate Professor, School of Aerospace, Tsinghua University

[3]  Professor, School of Aerospace, Tsinghua University

# 1. Introduction

The problem of the third-body perturbation has been studied for many years. Plenty of papers contributed research, the details of which were well reviewed in (Broucke 2003; Prado 2003). In the past century, the method of averaging technique to deal with the third-body perturbation already aroused great attentions (Musen et al. 1961; Cook 1962; Lorell 1965; Harrington 1969; Williams and Benson 1971; Lidov and Ziglin 1974; Ash 1976; Collins and Cefola 1979; Hough 1981; Šidlichovský 1983; Kwok 1985; Lane 1989; Kinoshita and Nakai 1991; Kwok 1991; Ferrer and Osacar 1994; Delhaise and Morbidelli 1997). Recently, the subject of averaged third-body perturbation is still a popular topic. A single averaged model over the short period of the satellite was used to study the effect of lunisolar perturbations on high-altitude Earth satellites (Solórzano and Prado 2007). The perturbing potential was doubly averaged first over one spacecraft orbit and second over one orbit of the third body (Scheeres 2001). The lunisolar effect that is double-averaged based on Lie transform was analyzed to deal with the resonance on a satellite's motion of the oblate Earth

(Radwan 2002). With the disturbing function expanded in Legendre polynomials at the second-order term, the double-averaged method was applied to analyze the long-term effect of a third body, and two important first integrals (energy and angular momentum) were used to discuss and classify properties of the perturbed orbits (Broucke 2003). An analytical double-averaged model with the expansion of Legendre polynomials up to the fourth-order term was developed, and a numerical study in the full elliptic restricted three-body model was made (Prado 2003). In the above papers, it was assumed that the perturbing body is in a circular orbit around the main body. Further, the analytical expansion to study the third-body perturbation was extended to the case where the perturbing body is in an elliptical orbit (Domingos et al. 2008).

The averaged models found wide applications in celestial mechanics, including analyzing long-term perturbations on lunar satellites (Folta and Quinn 2006; Carvalho et al. 2009a; Winter et al. 2009; Carvalho et al. 2010b), design of high altitude lunar constellations (Ely 2005; Ely and Lieb 2006), missions to Europa (Paskowitz and Scheeres 2005; Paskowitz and Scheeres 2006a; Paskowitz and Scheeres 2006b; Carvalho et al. 2010a), and missions to Ganymede and several other planetary moons (Russell and Brinckerhoff 2009).

In most above-mentioned papers, the *xy*-plane was defined as the orbital plane of the perturbing body instead of the equatorial plane of the main body, in which way the mean motion could be expressed in a simple form when only the third-body perturbation is concerned. However, when the gravitational non-sphericity effect is present or other perturbations are considered, the averaged models for third-body perturbations used in the above-mentioned papers may need redundant spherical trigonometric manipulations (Kinoshita and Nakai 1991; Yokoyama 1999) in order to be added into the spacecraft's equations of motion. To solve the problem, some papers (Carvalho et al. 2008; Carvalho et al. 2009a; Carvalho et al. 2009b; Winter et al. 2009; Carvalho 2010b; Carvalho 2011; Lara 2010; Lara 2011; Lara et al. 2009) neglected the inclination of the main body's equatorial plane with respect to its orbital plane (also known as obliquity or axial tilt); Other papers used the so-called Earth Orbit Frame to express the effect of the perturbing body in Earth-Moon system conveniently (Folta and Quinn 2006; Ely 2005; Ely and Lieb 2006).

However, in solar system, there exist some celestial bodies with non-ignorable obliquities. Table 1 lists the values of obliquities for different celestial bodies: Moon, Mars, Earth, Saturn, Uranus, Neptune, Pluto, Iapetus, Phoebe, and Nereid. The values of obliquities for these celestial bodies could not be considered as zero. The previous literature pointed out that the obliquity is probably the key to provide some important variations in eccentricities, and determines the extension of the chaotic zone (Yokoyama 1999).

**Table 1** Obliquities for different celestial bodies

| Celestial Bodies | Moon | Mars | Earth | Saturn | Uranus | Neptune | Pluto |
|---|---|---|---|---|---|---|---|
| Obliquity (deg) | 6.68 | 25.19 | 23.44 | 26.73 | 97.77 | 28.32 | 122.53 |

**Table 1**—Continued

| Celestial Bodies | Iapetus | Phoebe | Nereid |
|---|---|---|---|
| Obliquity (deg) | 15.215 | 26.723 | 30.011 |

This study extends the investigations of the double-averaged analytical model done by (Broucke 2003; Prado 2003; Domingos et al. 2008), and includes the action of the perturbing body's inclination, where the *xy*-plane is defined as the main body's equatorial plane. This paper is organized as follows. In Section 2, the dynamical model of the system is established. A double-averaged algorithm with respect to the period of the spacecraft and the period of the perturbing body is developed. In Section 3, this algorithm is applied to the Earth-Moon system and the Mars-Sun system. The full elliptic restricted three-body problem is also considered to verify the efficiency of the double-averaged algorithm. In Subsection 3.1, comparisons with the previous study for a lunar spacecraft are given in order to show the effect of the perturbing body's inclination. In Subsection 3.2, the application of the double-averaged algorithm to the Mars-Sun system is demonstrated and proved effective. The algorithm presented in this paper is also applicable to other celestial systems. Using this double-averaged model, the action of the third-body perturbation can be conveniently added to spacecraft's equations of motion when multiple perturbations are present.

## 2. Dynamical model

The system considered in this paper consists of three bodies: a main body with mass $m_0$, a massless spacecraft $m$, and a perturbing body (also known as a third body) with mass $m'$. All three bodies are assumed to be point masses. The reference frame $Oxyz$ is established with the origin $O$ located at the center of the main body, the $xy$-plane coinciding with the equatorial plane of the main body, the $x$-axis along the intersection line between the equatorial plane of the main body and the orbital plane of the third body, and the $z$-axis along the north pole of the main body. It is assumed that the perturbing body is in an elliptic inclined three-dimensional Keplerian orbit around the main body with semimajor axis $a'$, eccentricity $e'$, inclination $i'$, argument of pericentre $\omega'$, right ascension of the ascending node $\Omega'$, and mean motion $n'$ (given by the expression $n'^2 a'^3 = G[m_0 + m']$, where $G$ is the gravitational constant). It is obvious that $\Omega'=0$ due to the definition of the reference frame. The spacecraft is in a three-dimensional Keplerian orbit with semimajor axis $a$, eccentricity $e$, inclination $i$, argument of pericentre $\omega$, right ascension of the ascending node $\Omega$, and mean motion $n$ (given by the expression $n^2 a^3 = Gm_0$) around the main body, and is perturbed by the third body. The illustration of the system is presented in Fig. 1. Note that although the central body is assumed to be the point mass in the simplified model, when the model is applied to actual celestial systems, the equatorial plane of the central body and the orbital plane of the third body is different, and the inclination of the third body with respect to the equatorial plane of the central body takes effect.

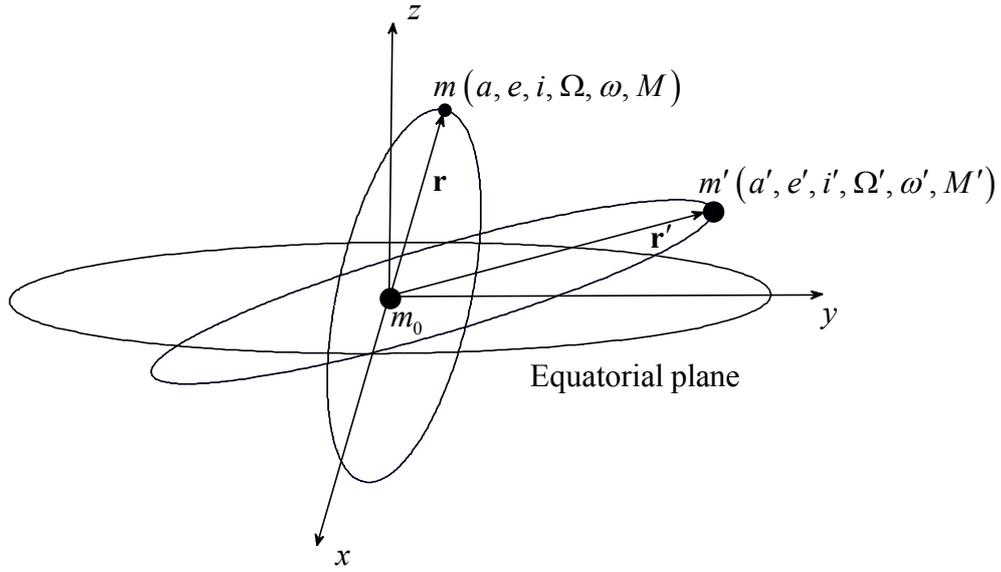

**Fig. 1** Illustration of the system.

Based on the theory of celestial mechanics (Murray and Dermott 1999), the disturbing function $R$ due to the action of the perturbing body can be obtained as

$$R = Gm' \left( \frac{1}{|\mathbf{r} - \mathbf{r}'|} - \frac{\mathbf{r}'}{r'^3} \cdot \mathbf{r} \right), \tag{1}$$

where $\mathbf{r}$ and $\mathbf{r}'$ are the orbital radius vectors of the spacecraft and the perturbing body, respectively. Further, the disturbing function $R$ can be expanded in the form of Legendre polynomials

$$R = \frac{\mu' G (m_0 + m')}{r'} \sum_{n=2}^{\infty} \left( \frac{r}{r'} \right)^n P_n(\cos\varphi), \tag{2}$$

where $\varphi$ is the angle between the radius vectors $\mathbf{r}$ and $\mathbf{r}'$, $P_n$ are the Legendre polynomials, and $\mu' = m'/(m_0 + m')$. Note that the $P_0(\cos\varphi)$ term is omitted because it is independent on orbital elements of the spacecraft, and the $P_1(\cos\varphi)$ term is eliminated after simple algebraic operations.

Assuming that the disturbing body is far from the spacecraft ($r \ll r'$), the disturbing function in the form of the Legendre polynomials expansion truncated up to the second order is shown as

$$R = \frac{\mu' G(m_0 + m')}{r'} \left(\frac{r}{r'}\right)^2 P_2(\cos\varphi)$$
$$= \frac{\mu' n'^2 a^2}{2} \left(\frac{a'}{r'}\right)^3 \left(\frac{r}{a}\right)^2 (3\cos^2\varphi - 1), \quad (3)$$

where the term $\cos\varphi$ is expressed as

$$\cos\varphi = \frac{\mathbf{r}}{r} \cdot \frac{\mathbf{r'}}{r'}. \quad (4)$$

Radius vectors **r** and **r'** can be expressed in terms of orbital elements of the spacecraft and the perturbing body. After some algebraic manipulations, Eq. (4) is arranged as

$$\cos\varphi = \alpha \cos f + \beta \sin f, \quad (5)$$

where $\alpha$ and $\beta$ are two intermediate variables,

$$\alpha = \cos\omega \cos\theta \cos u' + \sin\omega \sin i \sin i' \sin u' + \cos\omega \cos i' \sin\theta \sin u'$$
$$- \sin\omega \cos i \sin\theta \cos u' + \sin\omega \cos i \cos i' \cos\theta \sin u',$$
$$\beta = -\sin\omega \cos\theta \cos u' + \cos\omega \sin i \sin i' \sin u' - \sin\omega \cos i' \sin\theta \sin u' \quad (6)$$
$$- \cos\omega \cos i \sin\theta \cos u' + \cos\omega \cos i \cos i' \cos\theta \sin u';$$

$\theta$ is the difference of the perturbing body's and spacecraft's arguments of the longitudes,

$$\theta = \Omega - \Omega';$$

$u'$ is the perturbing body's argument of latitude,

$$u' = \omega' + f';$$

and $f$ and $f'$ are the true anomalies of the spacecraft and the perturbing body, respectively.

Substituting Eq. (5) into Eq. (3), the disturbing function is rewritten as

$$R = \frac{\mu' n'^2 a^2}{2} \left(\frac{a'}{r'}\right)^3 \left(\frac{r}{a}\right)^2 \left[3\alpha^2 \cos^2 f + 6\alpha\beta \sin f \cos f + 3\beta^2 \sin^2 f - 1\right]. \quad (7)$$

In order to reduce degrees of freedom of the system and remove the short-period terms, the averaging technique with respect to the mean anomaly of the spacecraft is applied, the definition of which is given as

$$\overline{K} = \frac{1}{2\pi} \int_0^{2\pi} K dM, \quad (8)$$

where $M$ is the mean anomaly that is proportional to time.

The double-averaged algorithm includes first averaging over the period of the spacecraft and second averaging with respect to the period of the perturbing body. The first average is performed over the short period of the spacecraft, and the disturbing function is derived as

$$\overline{R} = \frac{3\mu' n'^2 a^2}{4} \left(\frac{a'}{r'}\right)^3 \left[\left(\alpha^2 + \beta^2 - 2/3\right) + e^2 \left(4\alpha^2 - \beta^2 - 1\right)\right]. \quad (9)$$

The procedure of the first average is similar to the previous research (Broucke 2003; Prado 2003; Domingos et al. 2008).

For the next step, the second average is performed over the period of the perturbing body. In this step, the orbital elements of the spacecraft are assumed constant during the averaging (Prado 2003). When adding the action of the perturbing body's inclination $i'$, the case is a bit complicated.

Based on the theory of celestial mechanics (Murray and Dermott 1999), it is easy to prove that

$$\overline{\left(\frac{a'}{r'}\right)^3} = \left(1 - e'^2\right)^{-3/2}, \quad \overline{\sin u'} = 0, \quad \overline{\cos u'} = 0. \quad (10)$$

According to the trigonometric identities

$$\sin^2 u' = (1-\cos 2u')/2, \quad \cos^2 u' = (1+\cos 2u')/2, \quad \sin u' \cos u' = \sin 2u'/2,$$

it can be obtained that

$$\overline{\sin^2 u'} = 1/2, \quad \overline{\cos^2 u'} = 1/2, \quad \overline{\sin u' \cos u'} = 0. \tag{11}$$

Based on Eqs. (6) and (11), the following averages can be deduced by some algebraic manipulations,

$$\overline{\alpha^2 + \beta^2} = \frac{1}{2}\left(\cos^2\theta + \sin^2 i \sin^2 i' + \cos^2 i' \sin^2\theta + \cos^2 i \sin^2\theta \right. \\ \left. + \cos^2 i \cos^2 i' \cos^2\theta + \sin i' \cos i' \sin i \cos i \cos\theta\right), \tag{12}$$

$$\overline{4\alpha^2 - \beta^2} = \frac{1}{2}e^2\left[\left(5\cos^2\omega - 1\right)\cos^2\theta + \left(5\sin^2\omega - 1\right)\sin^2 i \sin^2 i' \right. \\ + \left(5\cos^2\omega - 1\right)\cos^2 i' \sin^2\theta + \left(5\sin^2\omega - 1\right)\cos^2 i \sin^2\theta \\ + \left(5\sin^2\omega - 1\right)\cos^2 i \cos^2 i' \cos^2\theta \\ - 5\sin\omega\cos\omega\sin\theta\cos\theta\cos i \sin^2 i' \\ + 5\sin\omega\cos\omega\sin i' \cos i' \sin\theta \sin i \\ \left. + \left(5\sin^2\omega - 1\right)\sin i' \cos i' \sin i \cos i \cos\theta\right]. \tag{13}$$

Finally, substituting Eqs. (10-13) into Eq. (9), the double-averaged disturbing function is shown as

$$\overline{\overline{R}} = \frac{3\mu' n'^2 a^2}{8}\left(1-e'^2\right)^{-3/2}\left\{\left(\cos^2\theta + \sin^2 i \sin^2 i' + \cos^2 i' \sin^2\theta\right.\right. \\ \left. + \cos^2 i \sin^2\theta + \cos^2 i \cos^2 i' \cos^2\theta + \sin i' \cos i' \sin i \cos i \cos\theta - \frac{4}{3}\right) \\ + e^2\left[\left(5\cos^2\omega - 1\right)\cos^2\theta + \left(5\sin^2\omega - 1\right)\sin^2 i \sin^2 i' \right. \\ + \left(5\cos^2\omega - 1\right)\cos^2 i' \sin^2\theta + \left(5\sin^2\omega - 1\right)\cos^2 i \sin^2\theta \\ + \left(5\sin^2\omega - 1\right)\cos^2 i \cos^2 i' \cos^2\theta - 5\sin\omega\cos\omega\sin\theta\cos\theta\cos i \sin^2 i' \\ + 5\sin\omega\cos\omega\sin i' \cos i' \sin\theta\sin i \\ \left.\left. + \left(5\sin^2\omega - 1\right)\sin i' \cos i' \sin i \cos i \cos\theta - 2\right]\right\}. \tag{14}$$

In order to derive the equations of motion, the partial derivatives of the disturbing function $R$ with respect to $a$, $e$, $i$, $\omega$, $\Omega$, and $M$ are required, and can be obtained as

$$\frac{\partial R}{\partial a} = 2C_1C_2 / a \{ (\cos^2\theta + \sin^2 i \sin^2 i' + \cos^2 i' \sin^2\theta$$
$$+ \cos^2 i \sin^2\theta + \cos^2 i \cos^2 i' \cos^2\theta + \sin i' \cos i' \sin i \cos i \cos\theta - \frac{4}{3} )$$
$$+ e^2 \big[ (5\cos^2\omega - 1)\cos^2\theta + (5\sin^2\omega - 1)\sin^2 i \sin^2 i'$$
$$+ (5\cos^2\omega - 1)\cos^2 i' \sin^2\theta + (5\sin^2\omega - 1)\cos^2 i \sin^2\theta \qquad (15)$$
$$+ (5\sin^2\omega - 1)\cos^2 i \cos^2 i' \cos^2\theta - 5\sin\omega\cos\omega\sin\theta\cos\theta\cos i \sin^2 i'$$
$$+ 5\sin\omega\cos\omega\sin i' \cos i' \sin\theta \sin i$$
$$+ (5\sin^2\omega - 1)\sin i' \cos i' \sin i \cos i \cos\theta - 2 \big] \},$$

$$\frac{\partial R}{\partial e} = 2eC_1C_2 \big[ (5\cos^2\omega - 1)\cos^2\theta + (5\sin^2\omega - 1)\sin^2 i \sin^2 i'$$
$$+ (5\cos^2\omega - 1)\cos^2 i' \sin^2\theta + (5\sin^2\omega - 1)\cos^2 i \sin^2\theta$$
$$+ (5\sin^2\omega - 1)\cos^2 i \cos^2 i' \cos^2\theta - 5\sin\omega\cos\omega\sin\theta\cos\theta\cos i \sin^2 i' \qquad (16)$$
$$+ 5\sin\omega\cos\omega\sin i' \cos i' \sin\theta \sin i$$
$$+ (5\sin^2\omega - 1)\sin i' \cos i' \sin i \cos i \cos\theta - 2 \big],$$

$$\frac{\partial R}{\partial \omega} = 5e^2 C_1 C_2 \sin 2\omega \big[ -\cos^2\theta + \sin^2 i \sin^2 i' - \cos^2 i' \sin^2\theta + \cos^2 i \sin^2\theta$$
$$+ \cos^2 i \cos^2 i' \cos^2\theta + \sin i' \cos i' \sin i \cos i \cos\theta \big] \qquad (17)$$
$$+ 5e^2 C_1 C_2 \cos 2\omega \big[ -\sin\theta\cos\theta\cos i \sin^2 i' + \sin i' \cos i' \sin\theta \sin i \big],$$

$$\frac{\partial R}{\partial i} = C_1 C_2 \{ \sin 2i (\sin^2 i' \cos^2\theta - \cos^2 i') + \cos 2i \sin i' \cos i' \cos\theta$$
$$+ e^2 \big[ \sin 2i (5\sin^2\omega - 1)(\sin^2 i' \cos^2\theta - \cos^2 i')$$
$$+ 5\sin i \sin\omega\cos\omega\sin\theta\cos\theta\sin^2 i' + 5\cos i \sin\omega\cos\omega\sin i' \cos i' \sin\theta \qquad (18)$$
$$+ \cos 2i (5\sin^2\omega - 1)\sin i' \cos i' \cos\theta \big] \},$$

$$\frac{\partial R}{\partial \Omega} = C_1 C_2 \{ -\sin 2\theta \sin^2 i \sin^2 i' - \sin\theta \sin i' \cos i' \sin i \cos i$$
$$+ e^2 \big[ \sin 2\theta \sin^2 i' (5\sin^2\omega\cos^2 i - 5\cos^2\omega + \sin^2 i)$$
$$- 5\cos 2\theta \sin\omega\cos\omega\sin\theta\cos i \sin^2 i' \qquad (19)$$
$$+ 5\cos\theta \sin\omega\cos\omega\sin i' \cos i' \sin i$$
$$- \sin\theta (5\sin^2\omega - 1)\sin i' \cos i' \sin i \cos i \big] \},$$

$$\frac{\partial R}{\partial M} = 0, \qquad (20)$$

where $C_1 = \frac{3}{8}\mu' n'^2 a^2$, $C_2 = (1-e'^2)^{-3/2}$.

The Lagrange's planetary equations for the variations of spacecraft's orbital elements are shown as (Chobotov 2002)

$$\frac{de}{dt} = \frac{1-e^2}{na^2 e} \frac{\partial R}{\partial M} - \frac{\sqrt{1-e^2}}{na^2 e} \frac{\partial R}{\partial \omega},$$

$$\frac{di}{dt} = \frac{\cot i}{na^2 \sqrt{1-e^2}} \frac{\partial R}{\partial \omega} - \frac{\csc i}{na^2 \sqrt{1-e^2}} \frac{\partial R}{\partial \Omega},$$

$$\frac{d\Omega}{dt} = \frac{1}{na^2 \sqrt{1-e^2} \sin i} \frac{\partial R}{\partial i},$$

$$\frac{d\omega}{dt} = \frac{\sqrt{1-e^2}}{na^2 e} \frac{\partial R}{\partial e} - \frac{\cot i}{na^2 \sqrt{1-e^2}} \frac{\partial R}{\partial i},$$

$$\frac{da}{dt} = \frac{2}{na} \frac{\partial R}{\partial M},$$

$$\frac{dM_0}{dt} = n - \frac{2}{na} \frac{\partial R}{\partial a} - \frac{1-e^2}{na^2 e} \frac{\partial R}{\partial e}.$$

(21)

Substituting Eqs. (15-20) into Eq. (21), the spacecraft's equations of motion can be derived.

Based on the equations of motion, some conclusions are obtained. It can be seen that the spacecraft's semimajor axis $a$ is constant during the motion. The time rates of the other five orbital elements are functions of $e$, $i$, $\omega$, and $\Omega$. It should be noticed that the spacecraft's $\Omega$ has an effect on the motion of the spacecraft in the double-averaged model and influences the other elements ($e$, $i$, $\omega$) due to the action of the perturbing body's inclination $i'$, which is different from the previous research (Scheeres 2001; Broucke 2003; Prado 2003; Solórzano and Prado 2007; Domingos et al. 2008). The mean anomaly $M$ does not appear in the right sides of the equations of motion because it is removed during by averaging.

Previous studies without considering the obliquity $i'$ show the role of the critical inclination 39° in the stability of near-circular orbits (Broucke 2003; Prado 2003;

Domingos et al. 2008). However, with an introduction of the obliquity's effect, the double-averaged model becomes complicated, and it is hardly to find a critical value for initial inclinations. The following section will indicate that the obliquity can cause slower changes in eccentricity. Thus, the obliquity could increase the stability of the spacecraft's orbit.

## 3. Numerical results

In this section, some numerical simulations are presented to show the validity of double-averaged model derived above. The results of the full elliptic restricted three-body problem are also given for the purpose of verification. The description of the full elliptic restricted three-body model can be seen in (Murray and Dermott 1999). The initial inclinations $i'$ of the perturbing body for these simulations are all nonzero. The scaling is made such that the semimajor axis of the perturbing body $a'$ is the unit of distance and $1/n'$ is the unit of time. Using this scaling, it can be found that the period of the perturbing body is $2\pi$.

**3.1 Application to the Earth-Moon system**

In order to compare with the previous study (Domingos et al. 2008), the Earth-Moon system is considered. The spacecraft is assumed to be in a three-dimensional orbit around Moon and its motion is perturbed by the disturbing body Earth. In this condition, one canonical time unit corresponds to 4.34 days. In

(Prado 2003), the initial conditions of the lunar satellite were $a_0$=0.01 (corresponding to 3844 km), $e_0$=0.01, and $\omega_0$=$\Omega_0$=0º with variable $i_0$, and the perturbing body Earth was assumed in a circular orbit. In (Domingos et al. 2008), the initial conditions of (Prado 2003) were used except that the perturbing body's eccentricity $e'$ was nonzero. The Earth-Moon system was generalized so that the eccentricity $e'$ varied in the range $0 \leq e' \leq 0.6$ and its effect was measured (Domingos et al. 2008).

For numerical simulations in this subsection, the initial conditions of spacecraft are taken the same as (Domingos et al. 2008), i.e. $a_0$=0.01, $e_0$=0.01, and $\omega_0$=$\Omega_0$=$M_0$=0º with variable $i_0$ from 55º to 80º. The action of the lunar obliquity $i'$=6.68º is considered additionally. In the following discussion, it can be seen that although the value of $i'$ is not large, its effect could not be neglected. It is assumed that the perturbing body's argument of pericentre $\omega' = 0°$. Besides, $a'$=1, $\Omega'$=0º, and $n'$=1. Only comparisons of the $e' = 0.1$ case with (Domingos et al. 2008) are presented here, and the other values of eccentricity $e'$ share similar behavior.

The evolutions of the spacecraft's inclination and eccentricity for different initial inclinations $i_0$ from 55º to 80º in order to compare the double-averaged model in this paper with (Domingos et al. 2008) are shown in Fig. 2. Seen from Fig. 2(a), it is found that the difference of evolutions of inclination between the results of the double-averaged model with considering $i'$ and the results based on Domingos et al. (2008) without considering $i'$ is becoming large with the increase of time. The previous literature indicated that the obliquity is probably the key to provide some important variations in eccentricities (Yokoyama 1999). Seen from Fig. 2(b), eccentricities can reach large values, which would make the orbits very elliptic. Note that eccentricities grow slower and their amplitudes are smaller with considering the lunar obliquity $i'$ than without considering $i'$, which confirms Yokoyama's speculation. Thus, the process of becoming elliptic can slow down due to the effect of the lunar obliquity, which would increase the stability of the spacecraft's orbit. In addition, it is evident that the results of both double-averaged models with and without considering $i'$ show smooth curves, just as expected. When considering the full elliptic restricted three-body model, the results are presented in Fig. 3. Compared Fig. 2 with Fig. 3, it can seen that the evolutions of $i$ and $e$ of the double-averaged model and the full elliptic restricted three-body model follow the similar trend, and show a good agreement.

The evolutions of argument of pericenter $\omega$ and right ascension of the ascending node $\Omega$ are shown in Fig. 4 and 5, respectively. It can be seen that $\omega$ keeps increasing while $\Omega$ keeps decreasing for different $i_0$. When the duration time is less than 400 canonical time units (corresponding to 1737 days), the evolutions of $\omega$ and $\Omega$ of the results of the double-averaged model with considering $i'$ and the full elliptic restricted three-body model are also well consistent with each other. However, the results based on Domingos et al. (2008) without considering $i'$ tend to deviate from the results of the full elliptic restricted three-body model with the time increasing.

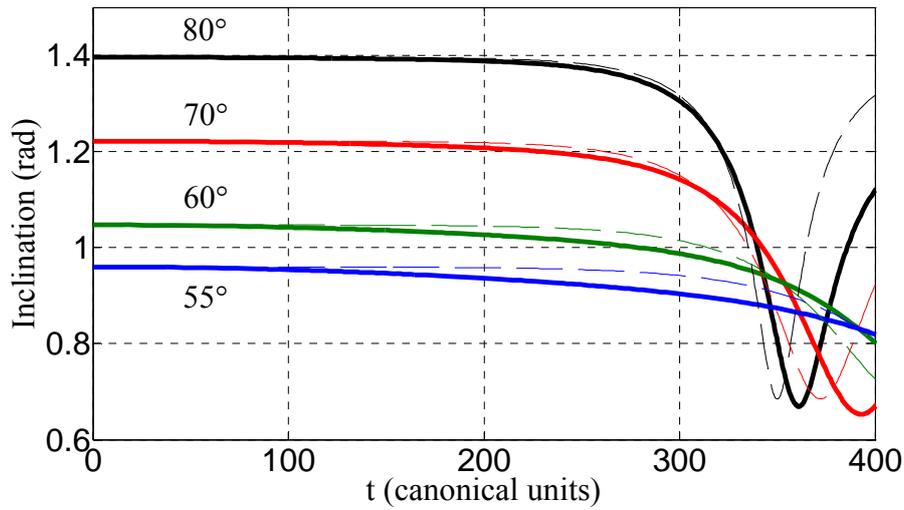

(**a**) Evolutions of inclination *i* over 400 canonical time units.

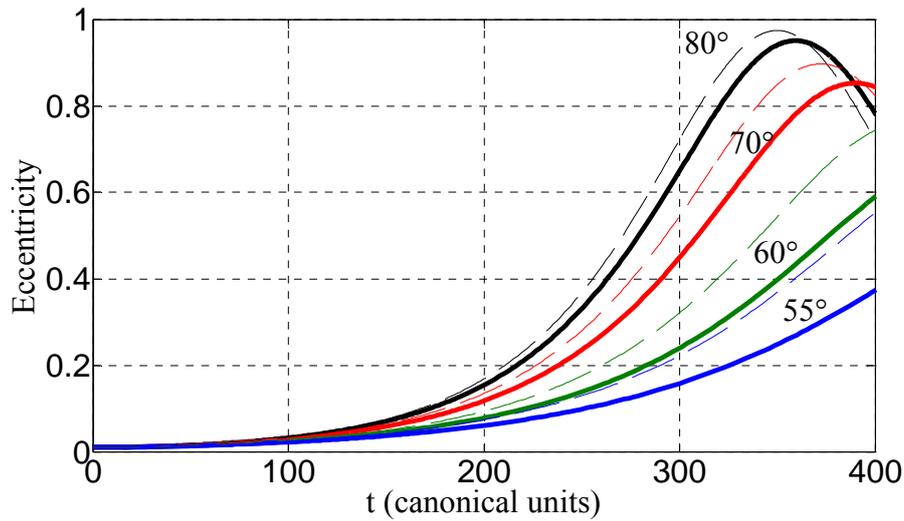

(**b**) Evolutions of eccentricity *e* over 400 canonical time units.

**Fig. 2** Evolutions of inclination and eccentricity for different spacecraft's initial inclinations over 400 canonical time units in the two different double-averaged models with and without considering $i'$. The perturbing body Earth is in an elliptic inclined orbit with $e' = 0.1$ and $i' = 6.68°$. Solid lines correspond to results including the action of the lunar obliquity $i'$, dashed lines to results based on Domingos et al. (2008) without considering $i'$.

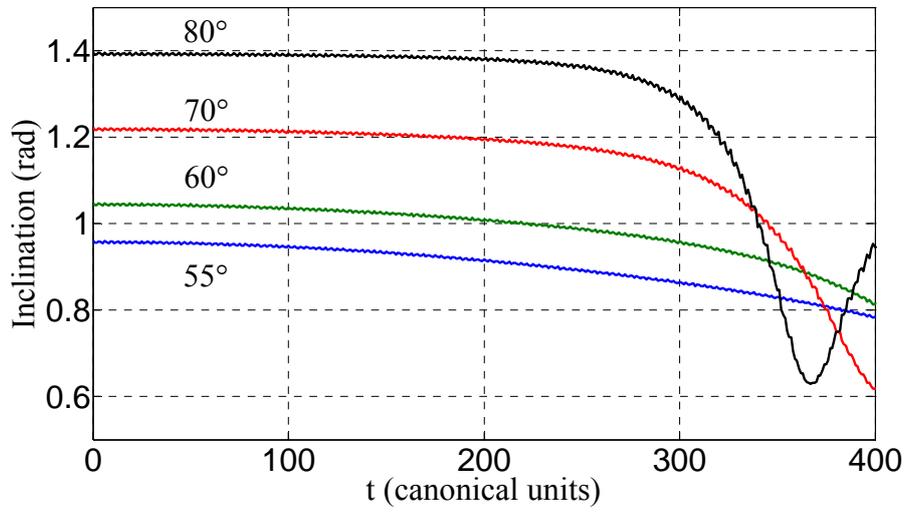

(**a**) Evolutions of inclination *i* over 400 canonical time units.

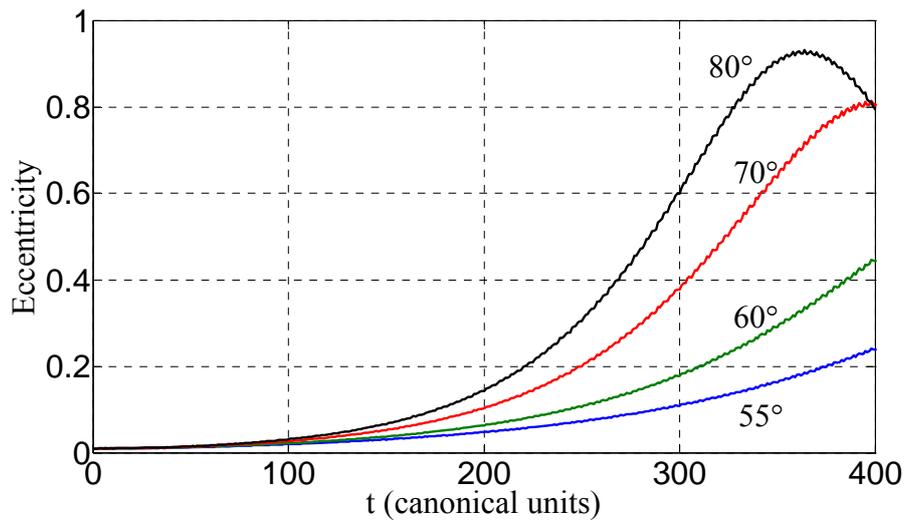

(**b**) Evolutions of eccentricity *e* over 400 canonical time units.

**Fig. 3** Evolutions of inclination and eccentricity for different spacecraft's initial inclinations over 400 canonical time units when considering the full elliptic restricted three-body model. The perturbing body Earth is in an elliptic inclined orbit with $e' = 0.1$ and $i' = 6.68°$.

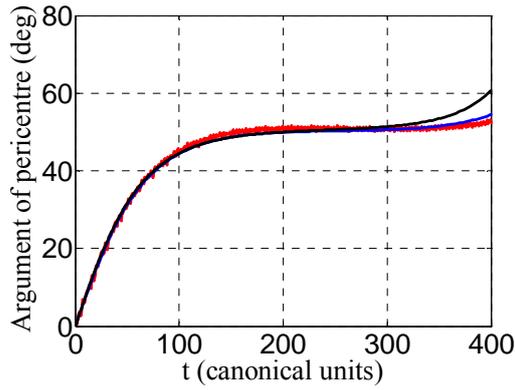
(a) Evolution of $\omega$ for $i_0 = 55°$

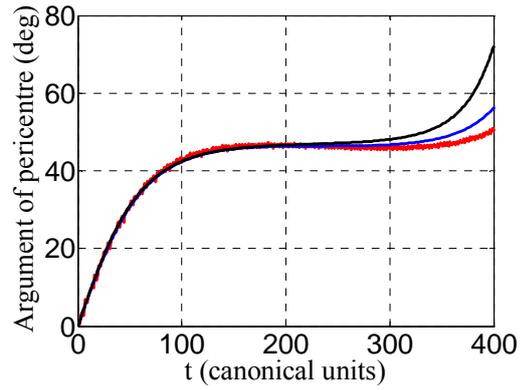
(b) Evolution of $\omega$ for $i_0 = 60°$

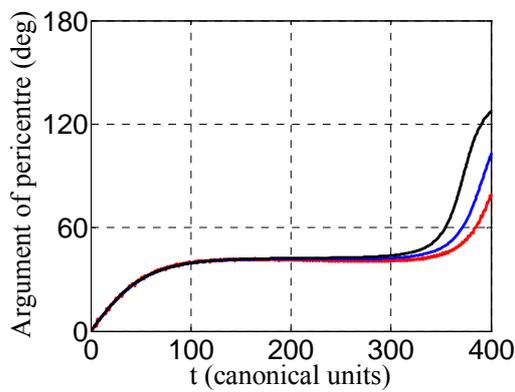
(c) Evolution of $\omega$ for $i_0 = 70°$

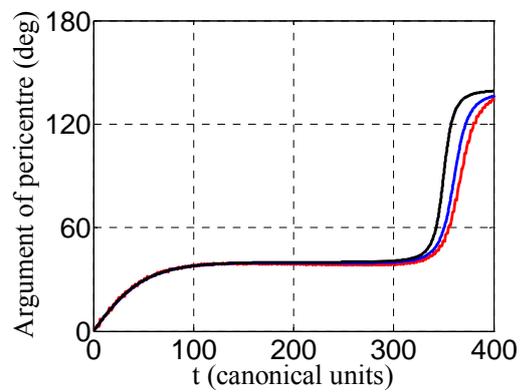
(d) Evolution of $\omega$ for $i_0 = 80°$

**Fig. 4** Evolutions of argument of pericenter $\omega$ for different spacecraft's initial inclinations over 400 canonical time units. The perturbing body Earth is in an elliptic inclined orbit with $e' = 0.1$ and $i' = 6.68°$. Lines in red correspond to results of the full elliptic restricted three-body model, lines in blue to results of the double-averaged model with considering the lunar obliquity $i'$, and lines in black to results based on Domingos et al. (2008) without considering $i'$.

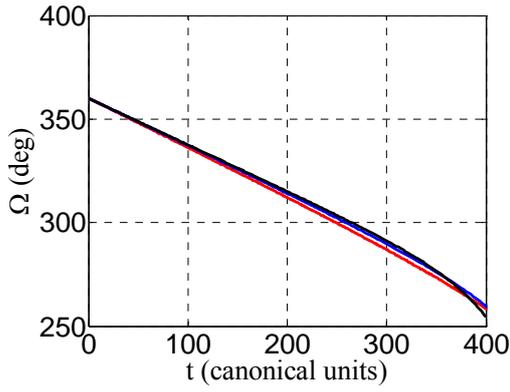

(a) Evolution of $\Omega$ for $i_0 = 55°$

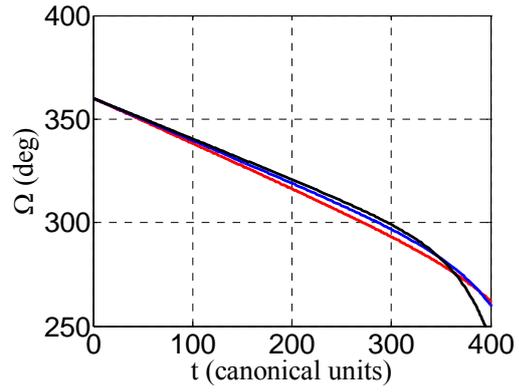

(b) Evolution of $\Omega$ for $i_0 = 60°$

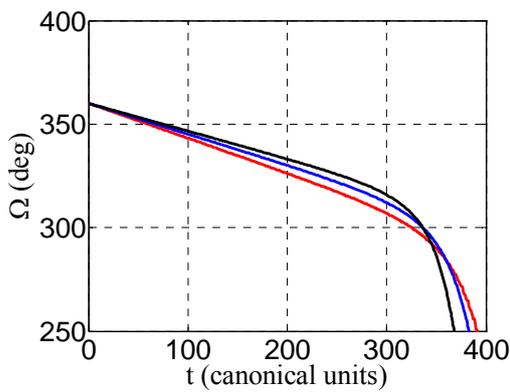

(c) Evolution of $\Omega$ for $i_0 = 70°$

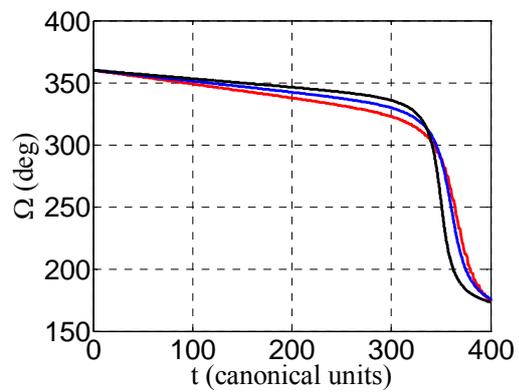

(d) Evolution of $\Omega$ for $i_0 = 80°$

**Fig. 5** Evolutions of right ascension of the ascending node $\Omega$ for different spacecraft's initial inclinations over 400 canonical time units. The perturbing body Earth is in an elliptic inclined orbit with $e' = 0.1$ and $i' = 6.68°$. Lines in red correspond to results of the full elliptic restricted three-body model, lines in blue to results of the double-averaged model with considering the lunar obliquity $i'$, and lines in black to results based on Domingos et al. (2008) without considering $i'$.

### 3.2 Application to the Mars-Sun system

Another case of the Mars-Sun system is also considered in order to show the validity of the double-averaged algorithm presented in Section 2. The values of the perturbing body's eccentricity $e'$ and inclination $i'$ are taken as the real values $e'=0.0935$, and $i'=25.19°$. It is assumed that the perturbing body's argument of pericentre $\omega' = 0°$. Besides, $a'=1$, $\Omega'=0°$, and $n'=1$. Using the scaling, it can be found that one canonical time unit corresponds to 109.32 days.

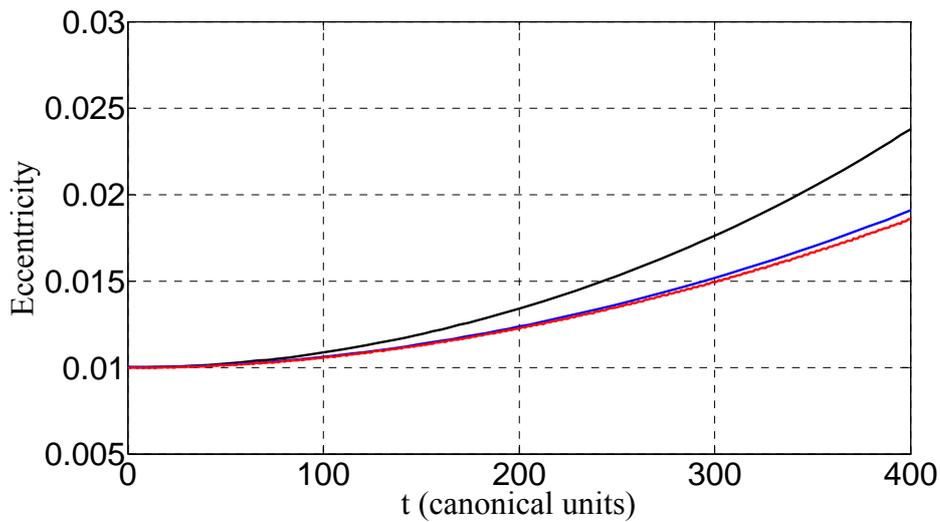

(**a**) Evolution of $e$ over 400 canonical time units.

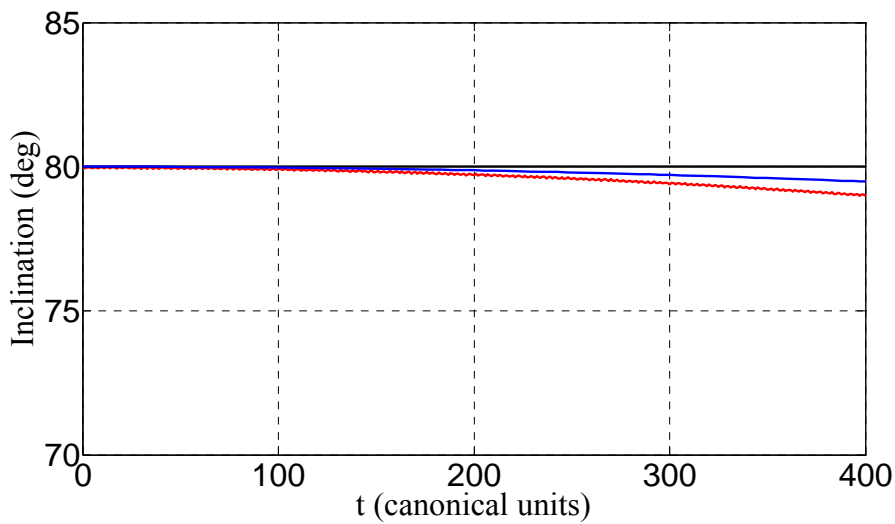

(**b**) Evolution of $i$ over 400 canonical time units.

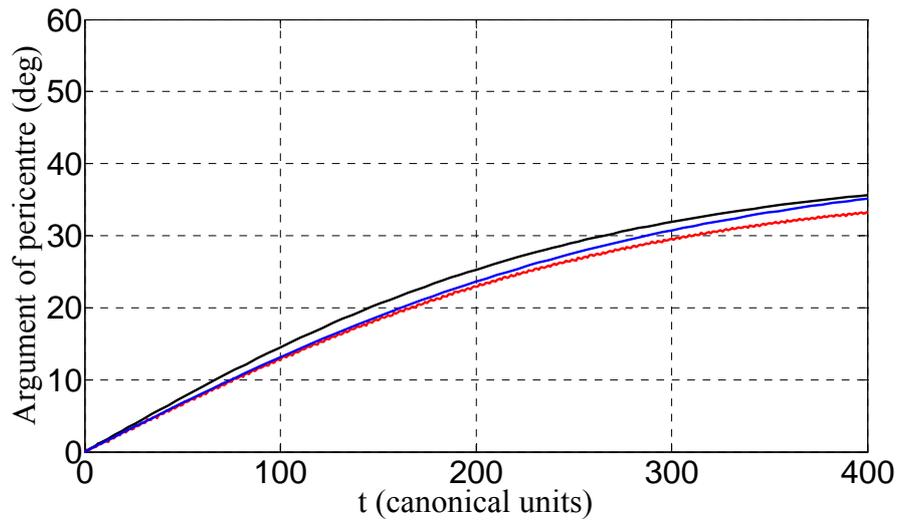

(**c**) Evolution of $\omega$ over 400 canonical time units.

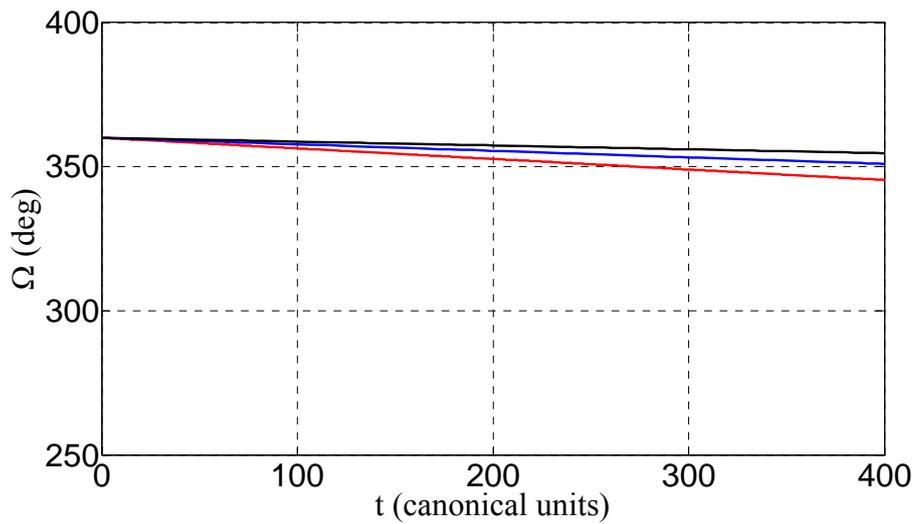

(**d**) Evolution of $\Omega$ over 400 canonical time units.

**Fig. 6** Evolutions of the spacecraft's orbital elements over 400 canonical time units. The perturbing body Sun is in an elliptic inclined orbit with $e' = 0.1$ and $i' = 25.19°$. Lines in red correspond to results of the full elliptic restricted three-body model, lines in blue to results of the double-averaged model with considering the Martian obliquity $i'$, and lines in black to results based on Domingos et al. (2008) without considering $i'$.

A near-circular Martian orbit is taken for instance. The initial orbital elements of the spacecraft are set as $a_0$=0.0001 (corresponding to 22792 km), $e_0$=0.01, $i_0$=80º, and $\omega_0$=$\Omega_0$=$M_0$=0º. The evolutions of the spacecraft's orbital elements $e$, $i$, $\omega$, and $\Omega$ are shown in Fig. 6. It is found that when the duration time is less than 400 canonical time units (corresponding to 43728 days), the evolutions of $e$, $i$, $\omega$, and $\Omega$ of the results of the double-averaged model with considering $i'$ and the full elliptic restricted three-body model follow the similar trend, and are well consistent with each other. It is also found that the difference between the results based on Domingos et al. (2008) without considering $i'$ and the full elliptic restricted three-body model tend to diverge with the time increasing.

Seen from Fig. 6, note that the evolution of $e$ is most affected by the Martian obliquity, which coincides with Yokoyama's speculation (Yokoyama 1999). Fig. 6(a) shows that the Martian obliquity cause much slower changes in eccentricity, which would increase the spacecraft's orbital stability. Within 400 canonical time units, $e$ and $\omega$ keep increasing, and $i$ and $\Omega$ keeps decreasing. Compared the results of the Earth-Moon system with those of the Mars-Sun system where the Sun is a distant disturbing body, it is found that the variation amplitudes of $e$, $\omega$, $i$, and $\Omega$ in the Mars-Sun system are much smaller than those in the Earth-Moon system. Thus, the near-circular spacecraft's orbits are more stable in the Mars-Sun system. From the above discussions, it is proved that the double-averaged model with considering $i'$ is effective to study long-term effect of solar perturbations for a high-altitude Martian satellite.

## 4. Conclusions

This paper develops a double-averaged analytical model with considering the obliquity to analyze the third-body perturbation. The efficiency of the double-averaged algorithm is verified in the full elliptic restricted three-body model, and it is proved that this algorithm is effective to predict the long-term motion for spacecrafts in both Earth-Moon system and the Mars-Sun system. The effect of the obliquity on the spacecraft's orbit is also explained. Comparisons with the previous study for a lunar satellite perturbed by Earth without considering obliquity show that the obliquity cause slower changes in eccentricity, which could increase the stability of the spacecraft's orbit. In the Mars-Sun system, it is also seen that the eccentricity grow slower due to the effect of the Martian obliquity. Besides, it is found that the variation amplitudes of orbital elements in the Mars-Sun system are much smaller than those in the Earth-Moon system, which would make the spacecraft's orbits more stable. The double-averaged algorithm presented in this paper is also applicable to other celestial systems.

## Acknowledgments

This work was supported by National Basic Research Program of China (973 Program) (2012CB720000) and the National Natural Science Foundation of China (No. 11072122).